\documentclass[twocolumn,preprintnumbers,amsmath,amssymb,nofootinbib,a4page]{revtex4}

\usepackage{graphicx}
\usepackage{dcolumn}
\usepackage{bm}
\usepackage{ams}
\usepackage{amssymb}
\usepackage{hyperref}

\def\be{\begin{equation}}
\def\ee{\end{equation}}
\def\e#1{\label{#1}\end{equation}}
\def\bea{\begin{eqnarray}}
\def\eea{\end{eqnarray}}
\def\ea#1{\label{#1}\end{eqnarray}}
\def\r#1{(\ref{#1})}
\def\bem#1{\begin{mathletters}\label{#1}}
\def\eml{\end{mathletters}}

\def\ket#1{{|#1\rangle}}
\def\bra#1{{\langle#1|}}

\def\mean#1{{\langle#1\rangle}}
\def\4#1{{\boldsymbol{#1}}}
\def\8#1{{\widetilde{#1}}}
\def\bse{\begin{subequations}}
\def\ese{\end{subequations}}
\def\eqref#1{(\ref{#1})}
\begin{document}


\title{Thermodynamical Control by Frequent Quantum Measurements\footnote{This is the presubmission version of
\mbox{\emph{Nature} \textbf{452}, 724}. The published version is available on the \emph{Nature} website: \url{http://dx.doi.org/10.1038/nature06873}.} }

\author{Noam Erez}
\author{Goren Gordon}
\author{Gershon Kurizki}
 \email{gershon.kurizki@weizmann.ac.il}
\affiliation{Department of Chemical Physics, Weizmann Institute of Science, Rehovot 76100, Israel.}

\author{Mathias Nest}
\affiliation{Theoretische Chemie, Universitaet Potsdam, Potsdam 14476, Germany.}

\begin{abstract}
Heat flow between a large ``bath'' and a smaller system brings
them progressively closer to thermal equilibrium while increasing
their entropy\cite{landau1980spe}. Deviations from this trend are
fluctuations involving a small fraction of a statistical ensemble
of systems interacting with the bath: in this respect, quantum and
classical thermodynamics are in
agreement\cite{landau1980spe,spo78,ali79,jar97,lin74}. Can there
be drastic differences between them? Here we address a distinctly quantum mechanical setting that displays
such differences: disturbances of thermal equilibrium between
two-level systems (TLS) and a bath\cite{gelman2003sdp} by frequent
and brief quantum
(non-demolishing)\cite{RevModPhys.75.715,braginsky1995qm,barnett1993tqt,haroche2006eqa}
measurements of the TLS energy-states. If the measurements are
frequent enough to induce either the Zeno or the anti-Zeno regime,
namely, the slowdown or speedup of the TLS
relaxation\cite{mis77,lan83,kof00,facchi2001po,kof04}, then the
resulting entropy and temperature of both the system and the bath
are found to be completely unrelated to what is expected by
standard thermodynamical rules that hold for memoryless
baths\cite{spo78,lin74}. The practical advantage of these
anomalies is the possibility of {\em very fast control} of heat
and entropy, allowing cooling and state-purification of quantum
systems much sooner than their thermal equilibration time.
\end{abstract}

\maketitle

To understand the origins of the predicted anomalies, consider a
thermal bath in equilibrium with an ensemble of quantum systems.
The energy of the quantum systems is briefly measured. How will these systems be affected? Classically, their equilibrium state
may remain intact, since measurements can be chosen to be
non-intrusive, i.e., involve no energy exchange, but merely
provide ``snapshots'' of the system. Likewise, quantum
mechanically, nearly-ideal (projective) measurements involve no
energy cost when performed by macroscopic detectors on isolated systems\cite{braginsky1995qm}. 
Yet finite-time coupling, followed by abrupt decoupling, of two quantum ensembles, which may be viewed as the detection of one ensemble by the other, may cause an increase of their mean total energy\cite{schulman06}. Here we address a different scenario pertaining to thermodynamics: a detector briefly (nearly-impulsively) measures only two-level systems (TLS) that are initially at thermal equilibrium with a much larger bath. We then ask: how will the temperature and entropy of these systems evolve both during and after the measurement via an interplay between the detector, the system and the bath?

Such nearly-impulsive quantum measurements in the basis of the system (energy) eigenstates,
chosen to be the ``pointer basis''\cite{RevModPhys.75.715} of the detector, must transfer energy, via the detector-system coupling, so as to
momentarily interrupt (override) the system-bath
interaction. This energy transfer, resulting in a change in system-bath
entanglement, triggers the distinctly quantum dynamics of both the system and the bath, which subsequently redistributes their mean energy and entropy in anomalous, unfamiliar, ways.

It is possible to use the information gained by such measurements to sort out
system subensembles according to their measured energy, in order
to extract work or entropy change, in the spirit of Maxwell's
demon\cite{scu01}. Here, however, we let the
entire TLS ensemble evolve regardless of the measured result, i.e., we
trace out the detector states. Our main concern is with the question: if we probe this evolution by a subsequent measurement,
how will the outcome depend on the time-separation of the two? The
answer should elucidate the virtually unknown short-time evolution of quantum
systems coupled to a bath.

Specifically, we analyze the scenario described above for a
TLS with energy separation $\hbar\omega_a$, that is
weakly coupled to a thermal bath of harmonic oscillators, characterized by a
correlation (memory) time of the bath response $t_c\gg1/\omega_a$, which typically marks the
onset of equilibrium. After equilibrium has been reached, we
perform $k=1,\ldots,K$ quantum non-demolition (QND)
measurements\cite{RevModPhys.75.715,braginsky1995qm,barnett1993tqt,haroche2006eqa}
of the TLS energy states at times separated by $\Delta
t_k=t_{k+1}-t_k$. Each measurement has a brief duration
$\tau_k\ll 1/\omega_a$. Our aim is to explore the evolution as a
function of time-separations between consecutive measurements in
the non-Markov domain, $\Delta t_k \lesssim 1/\omega_a \ll t_c$.
Such measurements do not resolve the energies of the TLS states,
due to the time-energy uncertainty. Yet they can discriminate
between states of different {\em symmetry}, e.g., different
angular momenta. In a TLS, the evolution of the mean energy or
state populations can be identified with the effective (spin)
temperature change\cite{coh92}, however rapidly it occurs.
In this uncharted domain, we show that consecutive brief
measurements entail several anomalies: (i) The quantum-mechanical non-commutativity of the system-detector and system-bath interactions causes the heat-up of the system at the expense of the detector-system coupling, but not at the expense of the coupling to the bath, only at very short $\Delta t_k$ compatible with the quantum Zeno effect (QZE)\cite{mis77}. (ii) A {\em transition from heating to cooling} of the TLS ensemble may occur as we vary the interval between consecutive measurements from $\Delta t_k\ll 1/\omega_a$ to 
$\Delta t_k\sim 1/\omega_a \ll t_c$. This marks the transition from $\Delta t_k$ compatible with the QZE to those compatible with the anti-Zeno effect (AZE)\cite{lan83,kof00,facchi2001po,kof04}. Remarkably, the cooling may occur even if the bath is initially hotter. (iii) Correspondingly,
{\em oscillations of the entropy} relative to that of the
equilibrium state take place, contrary to the Markovian notion of the second law of thermodynamics\cite{spo78,lin74}. 

This scenario is governed by the following total Hamiltonian of
the system that interacts with the bath and is intermittently
perturbed by the coupling of the system to the detector (measuring
apparatus):
\be
\label{H_total}
H(t)=H_{tot}+H_{SD}(t),\quad H_{tot}=H_S+H_B+H_{SB}.
\ee
Here $H_S$ is the Hamiltonian of the TLS, with ground and excited
states $\ket{g}, \ket{e}$, respectively; $H_B$ is that of the
thermal-bath composed of harmonic oscillators with energies
$\hbar\omega_\lambda$; $H_{SB}=\mathcal{S}\mathcal{B}$ is the
system-bath interaction Hamiltonian\cite{coh92} (the spin-boson interaction): a product of the
system-dipole (or spin-flip) operator $\mathcal{S}$ and the operator
$\mathcal{B}$ describing the bath excitations and deexcitations ; and $H_{SD}(t)$ is the
time-dependent measurement Hamiltonian that couples the system to a detector comprised of energy-degenerate ancillae (for details see Supplement A). In the coupling Hamiltonians
($H_{SB}$, $H_{SD}$) we {\em do not} invoke the rotating-wave
approximation (RWA)\cite{coh92}, namely, we do not impose energy
conservation between the system and the bath or the detector, on the time scales
considered\cite{kof04}.

The near-equilibrium state, $\rho_{tot}$, prior to a measurement
has several pertinent characteristics (Supplement B): (a) It
displays system-bath entanglement with off-diagonal matrix
elements $\langle e| \rho_{tot}| g \rangle \neq 0$. (b) The system
is described by a \emph{diagonal} reduced density matrix,
$\rho_S=Tr_B\rho_{tot}$, in the $H_S$ eigenbasis. (c) The mean
interaction energy $\langle H_{SB} \rangle$ is \emph{negative},
assuming $\rho_{tot}$ weakly deviates from the ground state of
$H_{tot}$: $\langle H_{SB}\rangle =\langle H_{tot}\rangle -
\langle H_S + H_B \rangle < 0$. This comes about since the
correction to the ground-state energy of $H_{tot}$ due to a
weakly-perturbing interaction $H_{SB}$ is negative (to the leading
second order).

We next consider the disturbance of this equilibrium state by a
nearly-impulsive (projective) quantum measurement
($\tau\rightarrow0$) of $S$, in the $|g\rangle,~|e\rangle$ basis. The measurement correlates the
TLS energy eigenstates with mutually orthogonal states of an
ancillary detector and the latter is then averaged (traced) over.
This measurement has distinctly quantum-mechanical consequences (Supplement A): it interrupts the
system-bath interaction\cite{kof00}, using the energy supplied by $H_{SD}(0<t<\tau)$ (the system-detector coupling) without changing $\mean{H_D}$. It thus eliminates the mean
system-bath interaction energy, whose pre-measurement value was
negative, as argued above:
\begin{equation}
\mean{H_{SB}(0)}<0\mapsto\mean{H_{SB}(\tau)}=0,\quad \mean{H_{SD}(t)}=-\mean{H_{SB}(t)}.
	\label{eq:H-SB-H_SD}
\end{equation}
We describe the detection process as a CNOT operation that retains the energy degeneracy of the detector $\mean{H_D}=0$, although its Von-Neumann entropy increases (Supplement A).

After the measurement (as $H_{SD}(t\ge\tau)=0$), time-energy uncertainty at $\Delta t \lesssim
1/\omega_a$ results in the breakdown of the RWA, i.e., $\langle
H_S + H_B \rangle$ is not conserved as $\Delta t$ grows. Only
$\langle H_{tot} \rangle$ is conserved, by unitarity, until the
next measurement. Hence, the post-measurement \emph{decrease} of
$\langle H_{SB} \rangle$ with $\Delta t$, signifying the
restoration of equilibrium: $\langle H_{SB}(\tau)\rangle =0
\rightarrow \langle H_{SB}(\tau+\Delta t)\rangle < 0$, is at the
expense of an \emph{increase} of $\langle H_S + H_B\rangle =
\langle H_{tot}\rangle - \langle H_{SB}\rangle$, i.e.,
\emph{heating} of the system and the bath, combined:
\be
\label{mean-energy}
\frac{d}{dt}\left[\mean{H_S}+\mean{H_B}\right]\Big|_{\tau+\Delta
t}>0, \quad \frac{d}{dt}\mean{H_{SB}}\Big|_{\tau+\Delta t}<0.
\ee

The post-measurement evolution of the system alone, described by
$\rho_S=Tr_B\rho_{tot}$, is not at all obvious. Its Taylor expansion holds at short
evolution times, $\Delta t \ll 1/\omega_a$,
\be
\label{rho-s-taylor}
\rho_S(\tau+\Delta t)\simeq \rho_S(\tau)+\Delta
t\dot\rho_S(\tau)+\frac{\Delta t^2}{2}\ddot{\rho}_S(\tau)+\ldots
\ee
The $0$th order term is {\em unchanged} by the measurement,
$\rho_S(\tau)=\rho_S(t\le0)$. The first derivative {\em vanishes}
at $t=\tau (\Delta t=0)$ due to the definite parity of the bath density-operator correlated to $\ket{g}$ or $\ket{e}$ (Supplement B). This initial post-measurement vanishing, $\dot{\rho}_S(\tau)=0$, is the QZE condition\cite{mis77,kof00,kof04,facchi2001po}. The
time evolution of $\rho_S$ is then governed by its second time
derivative $\ddot{\rho}_{S}(\tau)$, which can be shown (Supplement B) to have the same
sign as $\sigma_z =|e\rangle \langle e |-|g\rangle \langle g |$,
the population difference operator of the TLS. Hence, the second
derivative in \eqref{rho-s-taylor} is \emph{positive} shortly after the
measurement, consistently with Eq. \eqref{mean-energy}, if there
is no initial population inversion of the system, i.e., for
non-negative temperature.

The evolution of $\rho_S$ at longer
times (in the regime of weak system-bath coupling) may be
approximately described (as verified by our {\em exact} numerical
simulations\cite{nest2003dqd}, Supplement C) by the second-order
non-Markovian master equation (ME) (Fig.~1 -- main panel). The ME for
$\rho_S$, on account of its diagonality, can be cast into the
following population rate equations\cite{kof04}, dropping the
subscript $S$ in what follows and setting the measurement time to
be $t=0$:
\bea
\dot\rho_{ee}(t) &=& -\dot\rho_{gg}(t) =
R_g(t)\rho_{gg}-R_e(t)\rho_{ee},\\
\label{R-def}
R_{e(g)}(t) &=&  2\pi t\int_{-\infty}^\infty d\omega
G_T(\omega){\rm sinc}\left[(\omega\mp\omega_a)t\right],\\
\label{GT-def}
G_T(\omega) &=&(n_T(\omega)+1)G_0(\omega)+n_T(-\omega)G_0(-\omega).
\eea
Here $sinc(x)=\frac{\sin(x)}{x}$, $G_T(\omega)$ is the
temperature-dependent coupling spectrum of the bath, $G_0(\omega)$
is the zero-temperature coupling spectrum with peak coupling
strength at $\omega_0$ and spectral width $\sim 1/t_c$ and
$n_T(\omega)=\frac{1}{e^{\beta \omega}-1}$ is the
inverse-temperature- ($\beta$-) dependent population of bath mode
$\omega$.

The entire dynamics is determined by $R_{e(g)}(t)$, the relaxation
rates of the excited (ground) states. Their non-Markov
time-dependence yields three distinct regimes: 

(i) At short times
$t\ll  1/\omega_a \ll t_c$ the $sinc$ function in (\ref{R-def}) is
much broader than $G_T$. The relaxation rates $R_e$ and $R_g$ are
then equal at any temperature, indicating the complete breakdown of the RWA
discussed above: $|g\rangle \rightarrow |e\rangle$ and $|e\rangle
\rightarrow |g\rangle$ transitions do not require quantum
absorption or emission by the bath, respectively. The rates
$R_{e(g)}$ then become \emph{linear} in time, manifesting the
QZE\cite{kof00,facchi2001po,kof04}:
\bea
\label{R-short}
&&R_{e(g)}(t\ll t_c)\approx2 \dot R_0 t,\\
\label{Rdot-def}
&&\dot{R}_0\equiv\int_{-\infty}^\infty d\omega G_T(\omega)  =
\mean{\mathcal{B}^2}.
\eea
This short-time regime
implies the {\em universal Zeno heating rate}:
\be
\frac{d}{dt}\left(\rho_{ee}-\rho_{gg}\right) \approx 4 \dot{R}_0 t(\rho_{gg}-\rho_{ee}).
\ee

(ii) At intermediate non-Markovian times, $t \sim 1/\omega_a$,
when the $sinc$ function and $G_T$ in \r{R-def} have comparable
widths, the relaxation rates $R_{e(g)}(t)$ exhibit several unusual
phenomena that stem from time-energy uncertainty. The change in the overlap of the $sinc$ and $G_T$ functions with time results
in damped aperiodic oscillations of $R_e(t)$ and $R_g(t)$, near
the frequencies $\omega_0-\omega_a$ and $\omega_0+\omega_a$,
respectively. This oscillatory time dependence that conforms
neither to QZE nor to the converse AZE of relaxation speedup\cite{lan83,kof00,facchi2001po}, will henceforth be dubbed the {\em
oscillatory Zeno effect} (OZE). Due to the negativity of the
$sinc$ function between its consecutive maxima, we can have a {\em negative relaxation rate}, 
which is completely forbidden by the RWA. Since $sinc\left[(\omega+\omega_a)t
\right]$ is much further shifted from the peak of $G_T(\omega)$ than
$sinc\left[(\omega -\omega_a)t \right]$, $R_g(t)$ is more likely
to be negative than $R_e(t)$ (Fig.~1(a), Fig.~2(a)). Hence,
$\rho_{gg}(t)$ may grow at the expense of $\rho_{ee}(t)$ more than allowed by the thermal-equilibrium detailed balance. This may cause {\em transient cooling}, as detailed below.

(iii) At long times
$t\gg t_c$, the relaxation rates attain their Golden-Rule (Markov)
values\cite{kof04}
\be
\label{R-long}
R_{e(g)}(t\gg t_c) \simeq 2\pi G_T(\pm\omega_a).
\ee
The populations then approach those of an equilibrium Gibbs state whose
temperature is equal to that of the thermal bath (Fig.~1 -- main panel).

We now turn to entropy dynamics. One may always
\emph{define} the entropy of $\rho_S$ \emph{relative} to its
equilibrium state $\rho_0$ (``entropy distance'') and the negative
of its rate of change, as\cite{ali79,lin74}:

\begin{subequations}
\be \bm{S}(\rho_S(t)||\rho_0)\equiv \mathrm{Tr}\{\rho_S(t) \ln
\rho_S(t)\}-\mathrm{Tr}\{\rho_S(t) \ln \rho_0\}
\ee
\be
\bm{\sigma}(t) \equiv -\frac{d}{dt} \bm{S}(\rho_S(t)||\rho_0).
\label{sigma-def}
\ee
\end{subequations}
\emph{Only in the Markovian realm}, $\bm{\sigma}$ thus defined is
identified as the ``entropy production
rate''\cite{spo78,ali79,lin74}, where $\bm{\sigma}(t) \geq 0$ is a
statement of the second law of thermodynamics in this realm. Since $\rho_S$ is
diagonal, it follows (Supplement D) that $\bm{\sigma(t)}$ is
positive iff
$\frac{d}{dt}\left|\rho_{ee}(t)-(\rho_0)_{ee}\right|\leq 0$,
consistently with the interpretation of the relative entropy
$\bm{S}(\rho_S||\rho_0)$ in (\ref{sigma-def}) as the entropic
``distance'' from equilibrium. Conversely, whenever the
oscillatory $\rho_{ee}(t)$ drifts away from its initial or final
equilibria, $\bm{\sigma}$ takes negative values (Fig.~1(b)).


\begin{figure}[ht]
\includegraphics[width= 8cm]{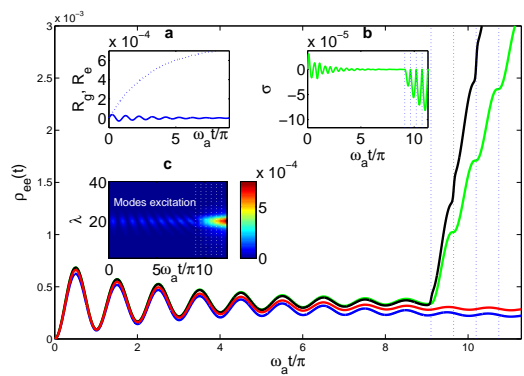}
\caption{\label{fig1} System and bath evolution as a function of time. Main
panel: Excited-level population as a function of time for
initially zero-temperature product state, followed by relaxation
to quasi-equilibrium and then subjected to a series of
measurements (vertical dashed lines). Measurements of finite
duration ($\tau_k=0.11/\omega_a$) (black line) results in somewhat
larger heat-up than impulsive measurements (red line), but the
dominant effect is the same for both. Observe the agreement
between $2^{nd}$ order master equation, two-quanta exchange
with a discrete bath (Suppl. A), and exact numerical solution for a discrete
bath of $40$ modes. (a) Relaxation rates, $R_g$, $R_e$ as a function of time.
(b) ${\bm\sigma}(t)$ (negative of relative entropy rate of change). 
(c) Excitations as a function of $t$ of the $40$ modes in the two-quanta model. 
Parameters: $t_c=10/\omega_a$, $\omega_0=\omega_a$,
$\gamma=0.07\omega_a$.}
\end{figure}


\begin{figure}[ht]
\includegraphics[width= 8cm]{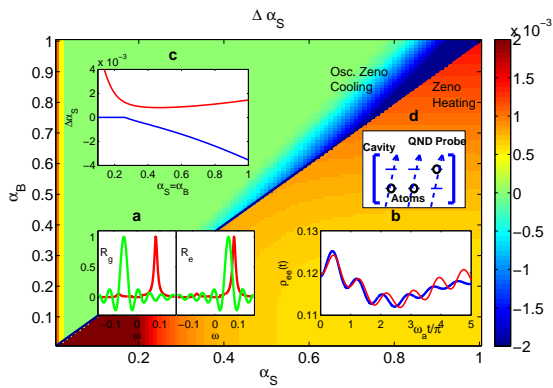}
\caption{\label{fig2} Maximal system heating and cooling. Main panel: Maximum
heating (lower half) and cooling (upper half) of the system, for
different
 system- ($x$-axis) and bath- ($y$-axis) initial temperatures:
 $\alpha_S\neq\alpha_B$,
 where $\alpha_{S(B)}= 1/\beta_{S(B)}\hbar\omega_a$. (a) $R_g(t)$ and $R_e(t)$ (Eq.~\eqref{R-def}) depicted as spectral overlaps of the relevant functions. (b) Example of a system experiencing first Zeno
 heating, then \emph{oscillatory}-Zeno cooling. (c) Maximal Zeno heating (red)
 and subsequent maximal cooling (blue) as a function of common initial temperature
 of system and bath. Note the {\em critical temperature} for
 oscillatory-Zeno cooling. Parameters: $t_c=10/\omega_a$,
$\omega_0=\omega_a/0.7$, $\gamma=4.36\omega_a$. These effects can be {\em strongly
magnified} by choosing other suitable parameters. (d) Possible experimental setup.}
\end{figure}


In order to realistically model the repeated measurements,
i.e. give them finite duration, we assume a smooth temporal
profile of the coupling to the detector (Supplement A). The $k$-th
measurement then occurs at time $t_k$ and has a duration of
$\tau_k$. Figure~1(main panel) compares the population evolution via
projective (impulsive) and finite-duration measurements with
$\tau_k\sim0.1/\omega_a$. Finite-duration measurements increase
the Zeno-heating as compared to impulsive ones due to the extra
energy supplied by the coupling to the detector. However, the
basic effect is seen (Fig.~1 -- main panel)  to be the same and is governed by
$\mean{H_{SB}}$ change in Eq.~\eqref{mean-energy}.
Counter-intuitively, finite-duration measurements are able to
\emph{increase the  cooling},
despite the extra energy coupled in by the apparatus.

If we repeat this procedure often enough, the TLS will either
increasingly heat up or cool down, upon choosing the time
intervals $\Delta t_k$ to coincide with either peaks or troughs of the
$\rho_{ee}$ oscillations, respectively. The minimal value of
$\bm{\sigma}$ can also be progressively lowered with each
measurement (Fig.~1(b)). Since consecutive measurements affect the
bath and the system differently, they may acquire different
temperatures, which then become the initial conditions for
subsequent QZE heating or OZE cooling. The results are shown in
Fig.~2 for both different (main panel) and common (Fig.~2(c))
temperatures of the system and the bath. Remarkably, the system may heat up solely
due to the QZE, although the {\em bath is colder}, or cool down solely
due to the OZE or AZE, although the {\em bath is hotter} (Fig. 2 -- main panel). The
bath may undergo changes in temperature and entropy too (Fig.~1(c)). 

One experimental realization of these effects can involve atoms or
molecules in a microwave cavity (Fig.~2(d)) with controllable
finite-temperature coupling spectrum $G_T(\omega)$ centered at
$\omega_0$. Measurements can be effected on such a TLS
ensemble with resonance frequency $\omega_a$ in the microwave
domain, at time intervals $\Delta t_k \sim 1/(\omega_0 \pm
\omega_a)$, by an optical QND probe\cite{braginsky1995qm} at
frequency $\omega_p \gg\omega_a, \omega_0$. The probe pulses
undergo different Kerr-nonlinear phase shifts $\Delta \phi_{e}$ or
$\Delta \phi_{g}$ depending on the different symmetries (e.g.,
angular momenta) of $|e\rangle$ and $|g\rangle$. The relative
abundance of $\Delta \phi_e$ and $\Delta \phi_g$ would then
reflect the ratio $\rho_{ee}(t_k)/\rho_{gg}(t_k)$. Such QND
probing may be performed with time-duration much shorter than
$\omega_a^{-1}$, i.e. $\omega_a \tau_k\ll 1$, without resolving
the energies of $|e\rangle$ and  $|g\rangle$.

Since non-selective measurements increase the Von-Neumann entropy
of the detector ancillae, their entropic price precludes a
``perpetuum mobile'', if closed-cycle operation is attempted. Yet,
if our ancillae are laser pulses, they are only used once and we
may progressively change the TLS ensemble thermodynamics by consecutive
pulses, disregarding their entropic or energetic price. The
practical advantage of the predicted anomalies is the possibility
of {\em very rapid control} of cooling and entropy, which may be
attained after several measurements at $t\geq\omega_a^{-1}$ and is
only limited by the measurement rate. By contrast, conventional
cooling requires much longer times, $t\gg t_c$, to
reach thermal equilibrium.

The present findings establish a new link between frequent quantum
measurements and nonequilibrium thermodynamical anomalies: heat
and entropy rates of change with the ``wrong'' sign, as compared
to their usual monotonic approach to equilibrium. These anomalies
are determined by the oscillatory or {\em negative} values of the
non-Markovian quantum relaxation rates at short times corresponding to large energy uncertainty. 
They reveal unfamiliar aspects of post-measurement quantum dynamics: AZE, which has been conceived as a means of \emph{enhancing or
accelerating} the initial-state
change\cite{lan83,kof00,facchi2001po,kof04}, here can
either \emph{restore} the equilibrium state or further depart from
it via cooling. These anomalies
underscore the often forgotten fact that the system and the bath
are \emph{inseparable} (entangled)\cite{gelman2003sdp,ste01},
even under weak-coupling conditions, a fact that has profound implications on their short-time dynamics.

These results prompt further studies of a hitherto unexplored
non-Markovian time domain where existing  formulations of the
second law of thermodynamics are inapplicable\cite{spo78,lin74},
and neither is the common notion that heat always flows from hotter to colder ensembles. This domain may necessitate an in-depth
scrutiny of quantum thermodynamical concepts. In particular, the
need for temporal ``coarse-graining'' of entropy should be
examined.

\begin{acknowledgments}
We acknowledge the support of ISF, GIF and EC (SCALA IP).
\end{acknowledgments}


\newpage
\begin{center}
\textbf{SUPPLEMENTARY INFORMATION}
\end{center}

\section*{Supplement A. Measuring the Energy of a Two-Level System Interacting with a Bath: Dynamical Description}

\subsection*{A.1 Hamiltonians and the measurement process}
We consider the following Hamiltonians:
\be
\label{H_total2}
H(t)=H_{tot}+H_{SD}(t),\quad H_{tot}=H_S+H_B+H_{SB}.
\ee
Here $H_{tot}$ pertains to the coupled system and bath and consists of:
\bea
\label{H-S}
&&H_S = \hbar\omega_a\ket{e}\bra{e}, \\
\label{H-B}
&&H_B = \hbar\sum_\lambda\omega_\lambda a_\lambda^\dagger a_\lambda,  \\
\label{H-I}
&&H_{SB} = \mathcal{SB}, \mathcal{S} = \ket{g}\bra{e}+\ket{e}\bra{g},~~
\nonumber \\&& \mathcal{B} = \hbar \sum_\lambda\left(\kappa_\lambda a_\lambda+\kappa_\lambda^* a_\lambda^\dagger\right),
\eea
where $\mathcal{S}$ and $\mathcal{B}$ are the system and bath
factors, respectively, in the system-bath interaction operator,
$a_\lambda(a^\dagger_\lambda)$ are the annihilation (creation)
operators, and $\kappa_\lambda$ is the matrix element of the weak
coupling to bath mode $\lambda$. 

The detector (ancilla) qubits have energy-degenerate states $|0\rangle_k, |1\rangle_k$ so that
we may set the detector Hamiltonian to be zero 
\be 
H_D=0.
\ee
The time-dependent system-detector coupling (to the $k$th detector) has the form
\begin{widetext}
\be \label{H-DS}
H_{SD}(t)=\sum_k H_{SD,k}=\sum_{k=1}^M h_k(t) \ket{e}\bra{e}
\left(\ket{0}_{kk}\bra{0} + \ket{1}_{kk}\bra{1}-\ket{0}_{kk}\bra{1}
-\ket{1}_{kk}\bra{0}\right).
\ee
\end{widetext}
where
\be
h_k(t)= \frac{\pi}{4\tau_k} \left({\rm
tanh}^2\left(\frac{t-t_k}{\tau_k}\right) - 1\right) \label{eq7}
\ee
is a smooth temporal profile of the system coupling to the detector qubits during 
the $k$-th measurement that occurs at time $t_k$ and has a duration of $\tau_k$.

This form of the single-measurement Hamiltonian $H_{SD,k}$ was chosen so that (dropping the index $k$ and
taking the measurement interval to be $[0,\tau]$ for simplicity):

\be
e^{-i\int_0^\tau dt H_{SD}(t)/\hbar} = U_C \label{eq8}.
\ee
where $U_C$ denotes to the CNOT operation (with the $k$th detector qubit, the target qubit, denoted by the subscript $D$):

\bea
\ket{g}\ket{0}_D &\mapsto& \ket{g}\ket{0}_D \nonumber \\
\ket{e}\ket{0}_D &\mapsto& \ket{e}\ket{1}_D \nonumber \\
\ket{g}\ket{1}_D &\mapsto& \ket{g}\ket{1}_D \nonumber \\
\ket{e}\ket{1}_D &\mapsto& \ket{e}\ket{0}_D. \label{eq10}
\eea
Since we take the initial state of the detector to be $\ket{0}$, only the first two rows play a role.
If the measurement duration $\tau$ is much shorter than the other time scales, then only $H_{SD}$ is non-negligible, and  the entire action of $H(t)$ during
this time is well approximated by the CNOT operator $U_C$. This becomes exact in the impulsive limit $\tau \rightarrow 0$. 

The measurement consists in letting the TLS interact with the detector (a degenerate TLS) via $H_{SD}$. The measurement
outcomes are averaged over (for nonselective measurements), by tracing out the detector degree of freedom. The total effect on the system density-operator
is:

\be
\rho_S \mapsto Tr_D \left\{U_C\rho_S \otimes \ket{0}_{DD}\bra{0}\right\} = \ket{e}\bra{e}\rho_S\ket{e}\bra{e}+
\ket{g}\bra{g}\rho_S\ket{g}\bra{g} 
\ee
i.e., the diagonal elements are unchanged, and the off-diagonals are erased.
Since the TLS is entangled with the bath, the effect of the measurement in Eqs. \eqref{eq8} - \eqref{eq10} is:
\bea
&& \rho_{tot} \mapsto Tr_D \left\{U_C\rho_{tot} \otimes \ket{0}_{DD}\bra{0} \right\} \\ 
&&  = \ket{e}\bra{e}\rho_{tot}\ket{e}\bra{e}+ \ket{g}\bra{g}\rho_{tot}\ket{g}\bra{g} \equiv \rho^B_{ee}\ket{e}\bra{e} + \rho^B_{gg}\ket{g}\bra{g}. \label{eq11}  \nonumber
\eea

\subsection*{A.2 Non-commutativity of $H_{SD}, H_{SB}$ and the measurement-induced vanishing of $\mean{H_{SB}}$ }

Here we demonstrate the validity of Eq. (2) of the main text, based on the post-measurement increase in $H_{tot}$ following the vanishing of $\mean{H_{SB}}$.
We show that this effect {\em disappears} if $H_{SD}(t)$ commutes with $H_{SB}$ and $H_S$ (and hence with $H_{tot}$).

For the commutative case, we have:
\bea
&& \mean{H_{tot}(\tau)} = Tr\left\{\rho_{SBD}(\tau)H_{tot}\right\} \\
&&=Tr \left\{ e^{-i \int_0^\tau dt [H_{tot} + H_{SD}(t)] } \left[\rho_{tot}(0)|0\rangle_D \phantom{\langle}_D\langle 0|\right]\times \nonumber \right. \\ &&\left. 
e^{+i \int_0^\tau dt' [H_{tot}+H_{SD}(t')]} H_{tot} \right\} \label{e34} \\
&& =Tr \left\{ \rho_{tot}(0)|0\rangle_D \phantom{\langle}_D\langle 0| e^{+i \int_0^\tau dt' [H_{tot}+H_{SD}(t')]} H_{tot} \right. \nonumber \\ &&\left. 
\times e^{-i \int_0^\tau dt [H_{tot} + H_{SD}(t)] } \right\} \label{e35} \\
&&=Tr\left\{\rho_{tot}(0)H_{tot}\right\}  \equiv  \mean{H_{tot}(0)}, \label{e36}
\eea
where $\rho_{SBD}$ is the state of the combined system, bath, and detector.
The cyclic property of the trace was used in \eqref{e35}, and the commutativity
of $H_{SD}$ and $H_{tot}$ in \eqref{e36}.

Compare this now to our {\em non-commutative} model (Eqs. \eqref{H-I} - \eqref{eq8})
\be
e^{-i\int_0^\tau dt H_{SD}(t)}\ket{0}_D = U_C|0\rangle_D = \ket{1}_D \ket{e}\bra{e} + \ket{0}_D \ket{g}\bra{g}.  \label{e37}
\ee
Since $H_{SD}$ in Eq. \eqref{H-DS} commutes with $H_S$, we may consider the evolution of $\mean{H_{SB}(\tau)}$, 
rather than $\mean{H_{tot}(\tau)}$.
In the impulsive limit ($\tau \rightarrow 0$), we can drop $H_{tot}$ in the exponent of Eq. \eqref{e35}, and then use the LHS of \eqref{e37} 
to obtain:
\be
\mean{H_{SB}(\tau)} = Tr \left\{ \rho_{tot}(0) \phantom{\rangle}_D\bra{0} U_{C}^\dagger H_{SB}(0) U_{C} \ket{0}_D  \right\}.
\ee
Finally, using the RHS of \eqref{e37} and \eqref{H-I}, we get:
\be
\phantom{\rangle}_D\bra{0} U_{C}^\dagger H_{SB}(0) U_{C} \ket{0}_D =0  \rightarrow  \mean{H_{SB}(\tau)} = 0.
\ee
This expresses the vanishing of $Tr \left \{\rho_{tot}(\tau)H_{SB}\right\}$ due to the diagonality of $\rho_{tot}(\tau)$ 
with respect to $S$ (Suppl. B). Since $H_D=0$, the detector mean energy is not affected by the CNOT action.

Hence, the measurement-induced interruption of the mean interaction energy, $\mean{H_{SB}(\tau)}=0$, and the resulting 
$\mean{H_S}+\mean{H_B}$ changes in Eq.(2) of the main text have a quantum mechanical origin: the non-commutativity of $H_{SB}$ and $H_{SD}$. 

\subsection*{A.3 Two-quanta approximation } 
\label{app-B}
Here we assume that the system and bath, governed by
Eqs.~\eqref{H_total2} - \eqref{H-DS} above, were in their respective
ground states prior to their interaction onset at $t=0$, followed
by a measurement at time $t_k$. As shown in Suppl. B, one should
allow for arbitrary excitations of the system and bath leading to
an infinite hierarchy of coupled equations for the populations of
$\ket{e}, \ket{g}$ and mode excitation numbers. Here, for
simplicity, we curtail this hierarchy, as is justified at short
times. The wave-function driven by $H(t)$ (Eq.~\eqref{H_total2})
acquires the following form, by allowing the system+bath to
receive or give away only $0$ or $2$ excitations (the lowest two
orders of the hierarchy expansion) through the coupling to the
detector:
\begin{widetext}
\bea
\ket{\psi(t)}&=&\sum_{l=0}^{2^M-1}\ket{\psi^{(l)}(t)}\ket{b_l}\\
\ket{\psi^{(l)}(t)}&=&
\8{\alpha}_{g,0}^{(l)}(t)\ket{g}\bigotimes_\lambda\ket{0}_\lambda
+\sum_\lambda
\8{\alpha}_{e,\lambda}^{(l)}(t)\ket{e}\ket{1}_\lambda\bigotimes_{\lambda'\neq\lambda}\ket{0}_{\lambda'}
\\&+&\sum_\lambda
\8{\alpha}_{g,\lambda}^{(l)}(t)\ket{g}\ket{2}_\lambda\bigotimes_{\lambda'\neq\lambda}\ket{0}_{\lambda'}
+\sum_{\lambda<\lambda'}
\8{\alpha}_{g,\lambda,\lambda'}^{(l)}(t)\ket{g}\ket{1}_\lambda\ket{1}_{\lambda'}\bigotimes_{\lambda''\neq\lambda,\lambda'}\ket{0}_{\lambda''}
\eea
\end{widetext}
where $b_l$ is the binary representation of $l$, labelling the
detector qubits. We transform to the frame where amplitudes are defined by:
\bea
&&\8\alpha_{g,0}^{(l)}=\alpha_{g,0}^{(l)}\\
&&\8\alpha_{e,\lambda}^{(l)}=e^{-i\omega_at-i\omega_\lambda
t}\alpha_{e,\lambda}^{(l)}\\
&&\8\alpha_{g,\lambda}^{(l)}=e^{-i2\omega_\lambda
t}\alpha_{g,\lambda}^{(l)}\\
&&\8\alpha_{g,\lambda,\lambda'}^{(l)}=e^{-i(\omega_\lambda+\omega_{\lambda'})
t}\alpha_{g,\lambda,\lambda'}^{(l)}.
\eea
Using the Schr\"{o}dinger equation, and integrating explicitly for
$\alpha_{g,0}^{(l)}$, $\alpha_{g,\lambda}^{(l)}$ and
$\alpha_{g,\lambda,\lambda'}^{(l)}$, we obtain the following
integro-differential matrix equation:
\bea
\label{alpha-dot}
\dot{\4\alpha}_{e}^{(l)}(t)&=&-\4R(t)\4\alpha_{e}^{(l)}(t)-i\4f^{(l)}(t)- \nonumber \\&& i\sum_{k=1}^Mh_k(t)\left(\4\alpha_e^{(l)}-\4\alpha_e^{(Q_k(l))}\right)\\
R_{\lambda,\lambda'}(t)&=&\int_0^tdt'\Big\{\kappa_{\lambda}^*\kappa_{\lambda'}
\left[e^{-i(\omega_{\lambda'}t'-\omega_\lambda t)}+
e^{-i(\omega_{\lambda'}t-\omega_\lambda
t')}\right]+\nonumber\\&&\delta_{\lambda,\lambda'}\sum_{\lambda''}\kappa_{\lambda''}^*\kappa_{\lambda''}
e^{-i\omega_{\lambda''}(t-t')}\Big\}e^{-i\omega_a(t'-t)}\\
\label{f-l}
f_{\lambda}^{(l)}(t)&=&
\kappa_\lambda^*e^{i\omega_at+i\omega_\lambda
t}\alpha_{g,0}^{(l)}(0)+\sqrt{2}\kappa_\lambda
e^{i\omega_at-i\omega_\lambda t}\alpha_{g,\lambda}^{(l)}(0)
+\nonumber\\&&\sum_{\lambda\neq\lambda'}\kappa_{\lambda'}e^{i\omega_at-i\omega_{\lambda'}
t}\alpha_{g,\lambda,\lambda'}^{(l)}(0)
\eea
where $\4\alpha_{e}^{(l)}(t)=\{\alpha_{e,\lambda}^{(l)}\}^T$, and
$Q_k(l)$ is the decimal representation of
$\{b_1,\ldots,1-b_k,\ldots,b_l\}$, describing the flipping of the
$k$th detector qubit. This flipping occurs within the $H_{SD}(t)$
activation interval defined in Eq.~\eqref{alpha-dot} by $h_k(t)$ (Eq.\eqref{eq7}).

Performing a brief measurement of the system at time $t_k$
(according to Eqs.~\eqref{alpha-dot}-\eqref{f-l}) ``splits'' the
subsequent evolution into two paths: (i) detection of the excited
state with probability
$P^{(e)}(t_k)=\sum_\lambda|\alpha_{e,\lambda}(t_k)|^2$, and
$\alpha_{g,0}^{(e)}(t_k+\epsilon)=\alpha_{g,\lambda}^{(e)}(t_k+\epsilon)=\alpha_{g,\lambda,\lambda'}^{(e)}(t_k+\epsilon)=0$,
where the superscript $(e)$ denotes the excited-state outcome;
(ii) detection of the ground state with probability
$P^{(g)}(t_k)=1-P^{(e)}(t_k)$ and
$\alpha_{e,\lambda}(t_k+\epsilon)^{(g)}=0$, the superscript $(g)$
denoting the ground-state. These subsequent independent evolutions
destroy the system-bath correlations, and can give rise to the
phenomena of Zeno heating and OZE cooling described in the text.

The plots in Fig.~1 of the text confirm the adequacy of the
present two-quanta approximation in describing the approach to
equilibrium and the measurement effects, compared to exact
numerical simulations (Suppl. C) or the second-order
master-equation approach.

\section*{Supplement B. Bath-System Entanglement Near Thermal Equilibrium: Pre- and Post-Measurement States}
\label{app-A}
The eigenstates of the total Hamiltonian $H_{tot}$
(Eq.~\eqref{H_total2}) exhibit entanglement between the system and
bath, due to the interaction term $H_{SB}$. The same is true of
thermal states ($Z^{-1}e^{-\beta H_{tot}}$), at least at low
temperatures and for weak coupling.

Likewise, if the system and bath are initially in a factorizable
eigenstate of $H_0=H_S+H_B$ and are subsequently exposed to the
total Hamiltonian, including the interaction term, they will
evolve into entangled system-bath states, at any temperature. If
the interaction is turned on adiabatically, an $H_0$ eigenstate
may be expected to evolve asymptotically into an eigenstate of
$H_{tot}$.

The aim of this Supplement is to prove the assertion that in all
these cases, the following properties of the joint state of the
system and bath obtain:

(i) $\rho_S$ is always diagonal (in the $H_S$ basis), which implies in turn $Tr \left\{\rho_{tot} H_{SB} \right\}=0$. 
(ii) The first moment of the bath excitation or deexcitation operator vanishes, causing the initial
vanishing of the time derivative of $\rho_S$ immediately after a measurement (the Zeno effect).
(iii) There is short-time post-measurement (Zeno) heating.

\subsection*{B.1 Pre-measurement evolution towards equilibrium}
We shall work in the interaction picture:
\begin{widetext}
\begin{eqnarray}
V_{I}(t)&\equiv& e^{+iH_{0}(t-t_{0})}H_{SB}e^{-iH_{0}(t-t_{0})}= \nonumber \\
(e^{-i\omega_{a}(t-t_{0})}|e\rangle\langle
g|&+&e^{+i\omega_{a}(t-t_{0})}|g\rangle\langle e|)
\sum_{k}(e^{i\omega_{0}(t-t_{0})}\kappa_{k}^{*}a_{k}^{\dagger}+\kappa_{k}a_{k}e^{-i\omega_{0}(t-t_{0})}), \\
|\Psi_{I}(t)\rangle &\equiv& e^{iH_{0}(t-t_{0})}|\Psi(t)\rangle=U(t,t_{0})|\Psi_{I}(t_{0})\rangle \label{e1a}, \\
U(t,t_{0})&=&1+\sum_{n=1}^{\infty}(-i)^{n}\int_{t_{0}}^{t}dt_{1}\int_{t_{0}}^{t_{1}}dt_{2}\cdots\int_{t_{0}}^{t_{n-1}}dt_{n}V_{I}(t_{1})V_{I}(t_{2})\cdots
V_{I}(t_{n})
\nonumber \\
&\equiv& \sum_{n=0}^\infty O_n(t). \label{e1b}
\end{eqnarray}
\end{widetext}
Let us denote the joint eigenstates of $H_B$ and $\hat{N}$ (the
total number operator) by:
\begin{equation}
|\bm{n}\rangle \equiv \mathcal{N}_{\bm{n}}\prod_j
(a_{k_j}^\dagger)^{n_j} |0\rangle,
\end{equation}
where $\bm{n}\equiv \{n_j\}_j$ and $\mathcal{N}_{\bm{n}}$ is the
appropriate normalization constant.


Consider first $|\Psi(t_0)\rangle=|\bm{n}\rangle\otimes|g\rangle$, where
$|{\bm{n}}\rangle$ has the $\hat{N}$ eigenvalue $n_{tot}$. We note
that $V_{I}(t)$(for any $t$) has the effect of flipping the $H_S$
state and transforming the bath state into a sum of states with
one more or one less excitation. Therefore the even terms
($O_{2m}(t)|\Psi(t_0)\rangle$) in the perturbation expansion of
$|\Psi_{I}(t)\rangle$, Eq.~\eqref{e1b} (counting the 1 as the zeroth term!), 
 are superpositions of states with excitation numbers
$n_{tot}+{\rm even}$, multiplied by $|g\rangle$, while the odd
ones ($O_{2m+1}(t)|\Psi(t_0)\rangle$) are superpositions of states
with $n_{tot}$+odd, multiplied by $|e\rangle$. Let us denote the
sum of the even terms of the series by $|B^{{\rm
even}}\rangle|e\rangle$ and that of the odd terms as $B^{{\rm
odd}}$, then:

\bea
|\Psi_{I}(t)\rangle&=&U(t,t_0)|\bm{n},g\rangle = |B^{{\rm
even}}_{\bm{n},g}(t)\rangle\otimes|g\rangle+|B^{{\rm
odd}}_{\bm{n},g}(t)\rangle\otimes|e\rangle \nonumber \\ &\equiv& |\Psi_{\bm{n},g}
(t)\rangle. \label{eq28}
\eea
Here $B^{{\rm even}}$ (respectively, $B^{{\rm odd}}$) is a sum of
$\hat{N}$-eigenstates with eigenvalues differing from $\hat{N}$ by
even (respectively, odd) numbers.

If the initial state is of the form
$|\Psi(t_0)\rangle=|\bm{n}\rangle\otimes|e\rangle$, the
time-evolved state is:

\bea
|\Psi_{I}(t)\rangle&=&U(t,t_0)|\bm{n},e\rangle = |B^{{\rm
even}}_{\bm{n},e}(t)\rangle\otimes|e\rangle+|B^{{\rm
odd}}_{\bm{n},e}(t)\rangle\otimes|g\rangle \nonumber \\ 
&\equiv& |\Psi_{\bm{n},e}(t)\rangle. \label{e26}
\eea

nd{equation}

Now consider the initial condition that the system and bath are in
a Gibbs state of $H_0$, with any inverse temperature $\beta$:

\begin{eqnarray}
&&\rho_{tot}(t_0)=Z_{tot}^{-1}e^{-\beta H_0} =   Z_S^{-1}e^{-\beta
H_S}  Z_B^{-1}e^{-\beta H_B} \label{eA8} \\ &&=Z_{tot}^{-1}
\sum_{\bm{n}} \left\{ e^{-\beta (\omega_g + \omega_{\bm{n}} )} |
\bm{n},g \rangle \langle \bm{n}, g | + e^{-\beta (\omega_e +
\omega_{\bm{n}} )} | \bm{n},e \rangle \langle \bm{n}, e |
 \right\} \nonumber
\end{eqnarray}
This initial state evolves at time $t\gg t_0$ into:
\bea
\rho_{tot}(t) =  Z_{tot}^{-1} \sum_{\bm{n}} \left\{ e^{-\beta
(\omega_g + \omega_{\bm{n}} )} | \Psi_{\bm{n},g}(t) \rangle
\langle \Psi_{\bm{n}, g}(t) |\right. \nonumber \\ \left. + e^{-\beta (\omega_e +
\omega_{\bm{n}} )} | \Psi_{\bm{n},e} \rangle \langle \Psi_{\bm{n},
e} |  \right\}  \label{eA10}
\eea
The resulting $\rho_{tot}$ has off-diagonal $\ket{e}\bra{g}$ and $\ket{g}\bra{e}$ elements by virtue of Eqs.
\eqref{eq28} and \eqref{e26}.  




Assuming that the adiabatic theorem can be applied (despite the
\emph{initial} degeneracy of the eigenstates), we have under
adiabatic switching on of $H_{SB}$:

\begin{equation}
\rho_{tot}(t\rightarrow\infty)=\sum_{\bm{n}}\sum_{m=g,e} e^{-\beta
E_{\bm{n},m}}|\Psi_{\bm{n},m}(t)\rangle \langle
\Psi_{\bm{n},m}(t) |, \label{eA11}
\end{equation}
$E_{\bm{n},m}$ being the eigenvalues of $H_0$.

This state has Gibbsian form, but with the original Boltzmann weights.
However, these weights are irrelevant for the parity of the state.

%

\subsection*{B.2 Diagonality of $\rho_S$ and post-measurement vanishing of its first order derivative}

We wish to establish the diagonality of $\rho_S$ before ($t\leq 0$) and after ($t\geq \tau$) the measurement, and the vanishing of $\dot{\rho}_S$ 
immediately after the measurement ($t=\tau$).
Due to the post-measurement vanishing of the off-diagonal elements of $\rho_{tot}$ (Eq. (\ref{eq11}), its derivative
immediately after the measurement, $\dot{\rho_S}(\tau)$, has the form: 	

\be
\dot{\rho_{S}}(\tau)=-i\left(e^{-i\omega_{a}\tau}|e\rangle\langle
g|-e^{+i\omega_{a}\tau}|g\rangle\langle e|\right)Tr_{B} \left\{
\mathcal{B}\left(\rho_{gg}^{B}-\rho_{ee}^{B}\right)\right\}.
\label{eq35}
\ee

For $\rho_{tot}(t) =  |\Psi_{\bm{n},g} (t)\rangle  \langle \Psi_{\bm{n},g} (t)|$ (Eq. \eqref{eq28}), we then have 

\bea
\left(\rho_S\right)_{eg}(t)&=&\bra{e}\rho_S(t)\ket{g}=Tr_{B}\langle e | \rho_{tot}(t)| g\rangle \nonumber \\ &=& \langle B^{{\rm
even}}_{\bm{n},e}(t)|B^{{\rm  odd}}_{\bm{n},e}(t)\rangle=0 \label{e24}
\eea
Hence, $\rho_S$ is diagonal at any time $t$. 

In addition, we have, by virtue of \eqref{eq35} at time $t=\tau$
\bea
\dot{\rho}_S(\tau) &\propto& Tr_{B} \left\{ \mathcal{B} \langle e (g) | \rho_{tot}| e
(g)\rangle \right\} \nonumber \\ &=& \langle B^{{\rm  even
(odd)}}_{\bm{n},e}(\tau)|\mathcal{B}|B^{{\rm  even
(odd)}}_{\bm{n},e}(\tau)\rangle=0 \label{e25}
\eea 

The same argument goes through upon permuting $e \leftrightarrow g$ everywhere for $\rho_{tot} =  |\Psi_{\bm{n},e} (t)\rangle  \langle \Psi_{\bm{n},e} (t)|$
(Eq.\eqref{e26}).

By linearity, using (\ref{e24}) and (\ref{e25}), the diagonality of $\rho_S(t)$ and the vanishing of $\dot{\rho_S}$ immediately
after the measurement are satisfied for (\ref{eA10}) and \eqref{eA11}.


\subsection*{B.3 Second derivative positivity at arbitrary time}


For the {\em factorizable} thermal state, 
\be
\rho_{tot}=Z^{-1}e^{-\beta H_0}=Z_B^{-1}e^{-\beta H_B} Z_S^{-1}e^{-\beta H_S}, 
\ee

we have:
\bea
&& \rho^B_{ee} \equiv \langle e | \rho_{tot}| e\rangle = \langle e
|Z_S^{-1}e^{-\beta H_S} | e \rangle Z_B^{-1}e^{-\beta H_B}  =
\left(\rho_S\right)_{ee} \rho_B \nonumber \\ &&~({\rm and~} e \leftrightarrow g ). \label{eq32}
\eea

For this $\rho_{tot}$, the second derivative of $\rho_S$ immediately after the measurement is (cf. Eq. \eqref{eq11})
\be
\ddot{\rho}_{S}(\tau)=2\sigma_z Tr_{B}\left\{
\mathcal{B}^2(\rho_{gg}^{B}-\rho_{ee}^{B})\right\}.
\label{eq14}
\ee
The scalar factor is positive:
\begin{equation}
Tr_{B}\left\{\hat{\mathcal{B}}^2\left(\rho^B_{gg}-\rho^B_{ee}\right)
\right\} = Tr_{B} \left\{\hat{\mathcal{B}}^2 \rho_B
\right\}\left( \left(\rho_S\right)_{gg}-\left(\rho_S\right)_{ee}\right) > 0, \label{eq47}
\end{equation} 
where we have used $Tr_B\{\rho^B_{gg(ee)}\}=\left(\rho_S\right)_{gg(ee)}$ which follows from the definition (Eq.\eqref{eq11}):
$\rho^B_{ee(gg)} = \bra{e(g)}\rho_{tot}\ket{e(g)}$.
The first factor in \eqref{eq47} is positive by virtue of the positivity of the operator $\hat{\mathcal{B}}^2$
($\hat{\mathcal{B}}$ being Hermitian), and the second is
positive iff there is no population inversion for the TLS. 

The combined (system- and bath-) equilibrium state satisfies: 
\be
\rho_{tot} = Z^{-1} e^{-\beta H_{tot}} = \rho_{tot}=Z^{-1}e^{-\beta \left( H_0 + O(H_{SB}^2) \right)}.
\ee 
Thus, for sufficiently weak coupling, Eq. \eqref{eq32} dominates.

We have made no use of the equality of the system and bath
temperatures. The argument goes through unchanged for
$\rho_{tot}=Z_B^{-1}e^{-\beta_B H_B} Z_S^{-1}e^{-\beta_S H_S}$.

\section*{C. Exact Numerical Simulations} The numerical
calculations have been done by the Multi-Configuration
Time-Dependent Hartree (MCTDH)
approach\cite{beck2000aag,nest2003dqd}. With this approach, very
large system-bath wave functions (with more than 100 degrees of
freedom) can be propagated in time with very high precision. In
order to treat finite temperatures we have sampled the Boltzmann
operator in an efficient way, by the random-phase thermal wave
function technique \cite{gelman2003sdp}.

\section*{D. Entropy Dynamics}
We may always \emph{define} the entropy of $\rho_S$
\emph{relative} to its equilibrium state $\rho_0$ and the negative
of its rate of change:
\bea \label{es1}
&&\bm{S}(\rho_S(t)||\rho_0)\equiv \mathrm{Tr}\{\rho_S(t) \ln
\rho_S(t)\}-\mathrm{Tr}\{\rho_S(t) \ln \rho_0\}
\label{RelEntDef}, \\
&&\bm{\sigma}(t) \equiv -\frac{d}{dt} \bm{S}(\rho_S(t)||\rho_0).
\label{sigma-def2}
\eea
\emph{Only in the Markovian realm}, $\bm{\sigma}$ thus defined is
identified as the `entropy production rate', i.e., the net rate of
change of the entropy of the system, after deducting the entropy
change due to the exchange of heat $\bm{Q}$ with the bath at
temperature $\bm{T}$ (in the absence of external work)
\cite{ali79}:
\be
\bm{\sigma}(t)=\frac{d\bm{S}(\rho_S)}{dt}-\frac{1}{\bm{T}}\frac{d\bm{Q}
}{dt}. \label{2a}
\ee where $\bm{S}(\rho_S)$ is the \emph{absolute} Von-Neumann entropy of the system.

It was proven by Lindblad \cite{lin74} that under any proper (completely positive)
quantum dynamical map, $\bm{M}$, the \emph{relative} entropy,
Eq.(\ref{RelEntDef}) cannot increase: $\bm{S}(\bm{M}\rho_S||\bm{M}
\rho_0) \leq \bm{S}(\rho_S||\rho_0)$. It then follows that under
Markovian evolution, which is described by such a map,
\be
\bm{\sigma}(t) \geq 0. \label{2b}
\ee
This is a statement of the second law of
thermodynamics\cite{spo78}.

The quantity $\bm{\sigma}(t)$ defined in Eq.~\eqref{sigma-def2}
(which only in the Markovian case is identified as the entropy
production rate of the TLS) has the following form, with
$(\rho_0)_{ee}$ as the long-time (equilibrium) excitation
probability
\be
\bm{\sigma}=-\dot{\rho}_{ee}\ln
\left[\frac{\rho_{ee}(1-(\rho_0)_{ee})}{(\rho_0)_{ee}(1-\rho_{ee})}\right].
\ee
Thus, $\bm{\sigma(t)}$ is positive iff
$\frac{d}{dt}\left|\rho_{ee}(t)-(\rho_0)_{ee}\right|\leq 0$,
consistently with the interpretation of the relative entropy
$\bm{S}(\rho_S||\rho_0)$ in (\ref{RelEntDef}) as a kind of
``distance'' from equilibrium. Conversely, whenever the
oscillatory $\rho_{ee}(t)$ drifts away from its initial or final
equilibria, $\bm{\sigma}$ must take negative values (Fig.~1(b)).

The condition for $\bm{\sigma} < 0$, amounts to:
\be
R_e(t)\rho_{ee}-R_g(t)\rho_{gg} < 0 {\rm~~and~~}
\ln\frac{\rho_{ee}}{\rho_{gg}}
\ee

Lindblad's theorem on the effect of quantum maps on the relative
entropy\cite{lin74} seems at first sight to contradict our
results. The resolution of this apparent paradox is that the
non-Markov evolution of $\rho_S(t)$ is not described by a map at
all! This can be seen (Figs.~1,2b) from the fact that the density
matrix oscillates and goes through the {\em same value} more than
once, with differing subsequent evolution. Hence,
$\bm{M}(0,t):\rho_{S}(0)\mapsto\rho_{S}(t)$ is a well defined
completely positive map, but it is not invertible: different
values of $\rho_{S}(0)$ may evolve into the same $\rho_{S}(t)$ for
some particular time $t$. This bears likeness to the findings in
Ref.~\cite{ste01} in a different context.



\bibliographystyle{unsrt}
\bibliography{Bibliography}


\end{document}